\documentclass[aps,superscriptadress,preprint]{revtex4}
\usepackage{graphicx}
\usepackage{amsmath}
\usepackage{mathrsfs}
\usepackage{amssymb}
\usepackage{booktabs}
\usepackage{tabularx}
\usepackage{textcomp}
\usepackage{feynmf}
\usepackage{color}
\begin{document}
\title{Baryon Number Fluctuations in Quasi-particle Model}

\author{Ameng Zhao$^{1}$}~\email[]{Email:zhaoameng@cxxy.seu.edu.cn}
\author{Xiaofeng Luo$^{2,3}$ }~\email[]{Email:xfluo@mail.ccnu.edu.cn}
\author{Hongshi Zong$^{4,5,6}$}~\email[]{Email:zonghs@nju.edu.cn}
\address{$^{1}$ Department of Foundation, Southeast University Chengxian College, Nanjing 210088, China}
\address{$^{2}$ Key Laboratory of Quark \& Lepton Physics (MOE) and Institute of Particle Physics, Central China Normal University, Wuhan 430079, China}
\address{$^{3}$Department of Physics and Astronomy, University of California, Los Angeles, California 90095, USA}
\address{$^{4}$ Department of Physics, Nanjing University, Nanjing 210093, China}
\address{$^{5}$ Joint Center for Particle, Nuclear Physics and Cosmology, Nanjing 210093, China}
\address{$^{6}$ State Key Laboratory of Theoretical Physics, Institute of Theoretical Physics, CAS, Beijing 100190, China}

\begin{abstract}

Baryon number fluctuations are sensitive to the QCD phase transition and QCD critical point. According to the Feynman rules of finite-temperature field theory, we calculated various order moments and cumulants of the baryon number distributions in the quasi-particle model of quark gluon plasma. Furthermore, we compared our results with the experimental data measured by the STAR experiment at RHIC. It is found that the experimental data can be well described by the model for the colliding energies above 30 GeV and show large discrepancies at low energies. It can put new constraint on qQGP model and also provide a baseline for the QCD critical point search in heavy-ion collisions at low energies.



\bigskip

\end{abstract}

\maketitle

\section{Moments of Net-Baryon Distributions and Quasi-particle Model of QGP}
Lattice QCD calculations indicate that at baryon chemical potential $\mu_B=0$, the transition from the quark-gluon plasma (QGP) to a hadron gas is a smooth crossover, while at large
$\mu_B$, the phase transition is of first order. The end point of the first order phase transition boundary is so called the critical point (CP). The fluctuations of net-proton number measured by the STAR experiment at RHIC suggest that the possible CP is unlikely below $\mu_B=200$ MeV {\cite{a1}}. Since the moments of the conserved quantities distributions, for example net-baryon number, in the relativistic heavy ion collisions are sensitive to the correlation length $\xi$ of the system \cite{a2}, and are believed to be good signatures of QCD phase transition and CP. Typically variances
($\sigma^2=<(N-<N>)^2>$) of the distributions are related to $\xi$ as $\sigma^2\sim\xi^2$. The numerators in skewness ($S=<(N-<N>)^3>/\sigma^3$) goes as $\xi^{4.5}$ and
kurtosis ($\kappa=<(N-<N>)^4>/\sigma^4-3$) goes as $\xi^7$.

On the other hand, the moments of baryon number are related to the various order baryon number susceptibilities \cite{a3}.  In order to cancel the volume, the products of the moments, $S\sigma$ and $\kappa\sigma^2$, are constructed as the experimental observables. The results in RHIC of these observables show a centrality and energy dependence \cite{a33},
which are not reproduced by a non-CP transport and hadron resonance gas model calculations. The deviations of $S\sigma$ and $\kappa\sigma^2$ below Skellam expectation are qualitatively consistent with a QCD based model which includes a CP \cite{a4}. The energy dependence of the $\kappa\sigma^2$ of net-proton distributions in Au+Au collisons show non-monotonic behavior, which is consistent with close to the CP \cite{a49}.

In this paper we apply the quasi-particle model (qQGP) of quark gluon plasma (QGP) to calculate the moments of net-baryon distributions. The qQGP model was first proposed by Peshier
et.al. \cite{a44} to study the non-ideal equation of state (EoS) by Lattice QCD results. Instead of real quarks and gluons with QCD interactions, the system is considered to be made up
of non-interacting quasi-quarks and quasi-gluons with thermal masses. Quasi-particles are thought to be quanta of plasma collective modes excited by
quarks and gluons through QCD interactions.

By now, some approaches have been proposed to study the qQGP model. The effective mass methods \cite{a44,a451,a452}, the approaches based on the Polyakov loop \cite{a453,a454,a455,a456,a457}, the approach based on Fermi liquids theory \cite{a458,a459} and so on. Comparing with the first and second approach, the third one is fundamentally different and powerful. Besides reproducing the EoS accurately, it is also successful in predicting the bulk and transport properties of QGP \cite{a458,a459}. 

 Gorenstein and Yang pointed out that initial quasi-particle model was
thermodynamically inconsistent and then reformulated the statistical mechanics (SM) to solve the inconsistency \cite{a45}.
But then the expressions of pressure and energy density are end up with an extra undetermined, temperature dependent terms, which need to be phenomenologically chosen. It should be paid attention that this reformulation in fact is based on mathematical identities
involving derivatives with respect to temperature and chemical potentials, used to redefine
average energy density and number density respectively. The qQGP model with reformulated SM by Gorenstein and Yang
has been studied by various groups \cite{a451,a448,a449,a410,a411,a412,a413}. On the other hand Bannur put forward another method which skip the thermodynamic  inconsistency by avoiding derivatives and instead use the original definition
of all thermodynamic quantities \cite{a5}. By doing this, the parameters of qQGP model are reduced. The results of qQGP model EoS, no matter which SM is adopted, are widely compared with Lattice data \cite{a449,a410,a411,a5,a50}. The results fit Lattice data well if the parameters are chosen properly.

Besides the EoS and the bulk and transport properties of QGP, quark-number susceptibilities are another important tool to test the reliability of qQGP model \cite{a71,a72}. The second order quark-number susceptibility of finite chemical potential and zero temperature \cite{a413} and of finite chemical potential and finite temperature \cite{a8,a81} are studied. But there are few works in qQGP model for the higher order susceptibilities associated with the results in RHIC so far. Since then, in this paper,we will calculated the moments of baryon distributions of proton and anti-proton in RHIC. By doing this, the study of qQGP model will be improved.

\section{Moments by Quasi-particle Model}
\label{two}

As mentioned in Ref. \cite{a45}, since the thermal mass of quasi-particle is temperature and chemical potential related, derivatives of the partition function with respect to temperature and chemical potentials destroy the thermodynamic consistence in the qQGP model. And then we have to redefine
average energy density and number density respectively by introducing  an extra undetermined, temperature dependent terms. Since the common method to obtain the susceptibilities of baryon number involve derivatives of the partition function with respect to baryon chemical potentials, then an extra term must be introduced in the calculation to maintain the thermodynamic consistence. To avoid it, we adopt the same way as Bannur has done. In Ref. \cite{a5}, Bannur gets the expectation of particle number by
\begin{equation}
<N>=\sum_{k}\frac{z\epsilon_ke^{-\beta\epsilon_k}}{1\mp z\epsilon_ke^{-\beta\epsilon_k}}\ ,
\label{1}
\end{equation}
instead of doing derivatives of the partition function
\begin{equation}
<N>=T\frac{\partial{Ln Z}}{\partial \mu}\ ,
\end{equation}
where $z$ is the fugacity, $\epsilon_k$ is the single particle energy and $Z$ is the partition function of particles (more detail can be found in Ref. \cite{a5}).
Similarly, in this paper we obtain the quark-number susceptibilities thermodynamic consistently by avoiding to make derivatives to the partition function. It should be emphasized that, rather than the method of Eq. (\ref{1}), we calculate the mathematical expectations of $<N>$ and $<N^n>$  directly by the field theory at finite temperature and chemical potential according the Lagrangian of quasi-quarks.

For the simplicity of calculation, we adopt the quasi-particle model of QGP here. In this model, the interaction of quarks and gluons is treated as an effective mass term \cite{a5}. The effective mass of quark is made up of the rest mass and the thermal mass,
\begin{equation}
m^2=m_{q0}^2+\sqrt{2}m_{q0}m_{th}+m_{th}^2\ ,
\label{mq}
\end{equation}
where $m_{q0}$ is the rest mass of up or down quark, and in this paper $m_{q0}=6.5$ MeV. The temperature and chemical potential dependent quark mass $m_{th}$ is
\begin{equation}
m_{th}^2(\mu,T)=\frac{g^2T^2}{18}\mathcal{N}_{f}(1+\frac{\mu^2}{\pi^2T^2})\ ,
\end{equation}
and $g^2$ is related to the two-loop order running coupling constant,
\begin{equation}
\alpha_s=\frac{6\pi}{(33-2\mathcal{N}_{f})\ln\frac{T}{\Lambda_T}\sqrt{1+a\frac{\mu^2}{T^2}}}(1-\frac{3(153-19\mathcal{N}_{f})}{(33-2\mathcal{N}_{f})^2}\frac{\ln(2\ln{\frac{T}{\Lambda_T}\sqrt{1+a\frac{\mu^2}{T^2}}})}{\ln{\frac{T}{\Lambda_T}\sqrt{1+a\frac{\mu^2}{T^2}}}})\ ,
\end{equation}
where $\alpha_s=g^2/4\pi$. In this paper, only up and down quarks are considered, so $\mathcal{N}_{f}=2$. The parameter $a$ mainly has two choice. One is equal to $(1.91/2.91)^2$ in the calculation of
 Schneider \cite{a6} and
the other is $(1/\pi)^2$ in a phenomological model of Letessier and Rafelski \cite{a7}.

The expectation of quark number is
\begin{equation}
<N_{q}>=<N_{u}>+<N_{d}>=\mathcal{N}_{c}\mathcal{N}_{f}<N>\ ,
\end{equation}
where $<N>$ is the quark number expectation of one single color and flavor. And the expectation of baryon number is $<N_{B}>=\frac{1}{3}<N_{q}>$. The variance of quark number is
\begin{eqnarray}
\begin{split}
<(N_{q}-<N_{q}>)^2>=&<[(N_{u}+N_{d})-<N_{u}+N_{d}>]^2>\\
=&<(N_{u}-<N_{u}>)^2>+<(N_{d}-<N_{d}>)^2>\\
&+2<(N_{u}-<N_{u}>)(N_{d}-<N_{d}>)>\\
=&<(N_{u}-<N_{u}>)^2>+<(N_{d}-<N_{d}>)^2>\\
=&{\mathcal{N}_{c}}^2\mathcal{N}_{f}<(N-<N>)^2>\ ,\\
\end{split}
\end{eqnarray}
since the up quarks and down quarks are independent, we have $<(N_{u}-<N_{u}>)(N_{d}-<N_{d}>)>=0$. Then the variance of baryon number is
\begin{eqnarray}
\begin{split}
\sigma^2 &=<(N_{B}-<N_{B}>)^2>\\
&=\frac{1}{{\mathcal{N}_{c}}^2}<(N_{q}-<N_{q}>)^2>=\mathcal{N}_{f}<(N-<N>)^2>\ ,\\
\end{split}
\end{eqnarray}
The skewness of baryon number is
\begin{equation}
S=\frac{<(N_{B}-<N_{B}>)^3>}{[\sigma^2]^{3/2}}=\frac{\mathcal{N}_{f}<(N-<N>)^3>}{[\mathcal{N}_{f}<(N-<N>)^2>]^{3/2}}\ ,
\end{equation}
and the kurtosis of baryon number is
\begin{eqnarray}
\begin{split}
\kappa &=\frac{<(N_{B}-<N_{B}>)^4>}{[\sigma^2]^2}-3\\
&=\frac{\mathcal{N}_{f}(<(N-<N>)^4>+3<(N-<N>)^2>^2)}{[\mathcal{N}_{f}<(N-<N>)^2>]^{2}}-3\\
&=\frac{\mathcal{N}_{f}(<(N-<N>)^4>-3<(N-<N>)^2>^2)}{[\mathcal{N}_{f}<(N-<N>)^2>]^{2}}\ ,\\
\end{split}
\end{eqnarray}
then the products of the moments constructed as the experimental observables, $S\sigma$ and $\kappa\sigma^2$, are
\begin{equation}
S\sigma=\frac{<(N-<N>)^3>}{<(N-<N>)^2>}\ ,
\label{s}
\end{equation}
\begin{equation}
\kappa\sigma^2=\frac{<(N-<N>)^4>-3<(N-<N>)^2>^2}{<(N-<N>)^2>}\ .
\label{kappa}
\end{equation}
Since the quarks are treated as the free quasi-particle with thermal masses, then it can be written as
\begin{equation}
-\beta(H-\mu N)=\int_{0}^{\beta}d\tau\int d^3x \overline{\psi}_{q}(-\gamma_{4}\frac{\partial}{\partial\tau}+i\overrightarrow{\gamma}\cdot\overrightarrow{\nabla}-m+\mu \gamma_{4})\psi_{q}\ ,
\end{equation}
where $\psi_{q}$ is the quark field and $\mu$ is the chemical potential of quarks ($\mu={\mu}_u={\mu}_d=\frac{1}{3}{\mu}_B$), then the quark number $N$ is
\begin{equation}
\beta N=\int_{0}^{\beta}d\tau\int d^3x \overline{\psi}_{q} \gamma_{4}\psi_{q}\ ,
\end{equation}
then the quark number expectation $<N>$ is
\begin{eqnarray}
\begin{split}
<N>&=\frac{\int\mathcal{D}\overline{\psi}_{q}\int\mathcal{D}\psi_{q}\int_{0}^{\beta}d\tau\int d^3x \overline{\psi}_{q}(\overrightarrow{x},\tau)\gamma_{4}\psi_{q}(\overrightarrow{x},\tau)\exp{(-\beta(H-\mu N))}}{\beta\int\mathcal{D}\overline{\psi}_{q}\int\mathcal{D}\psi_{q}\exp{(-\beta(H-\mu N))}}\\
&=-VT\sum_{k=-\infty}^{+\infty}\int\frac{d^3p}{(2\pi)^3}Tr[G(\widetilde{p}_k)\gamma_4]\\
&=VT\sum_{k=-\infty}^{+\infty}\int\frac{d^3p}{(2\pi)^3}\frac{4i\widetilde{\omega}_{k}}{\overrightarrow{p}^2+m^2+\widetilde{\omega}^2_{k}}\ ,\\
\label{N}
\end{split}
\end{eqnarray}
and this expression of quark-number is the same as that widely used in other works \cite{a47,a48}, where $\widetilde{p}_k=(\vec{p},\widetilde{\omega}_k)=(\vec{p},i\mu+\omega_k)$, $\omega_k=(2k+1)\pi T$. $\overline{\psi}_{q} \gamma_{4}\psi_{q}$ can be analogized as the interaction term, then the Feynman rules are \cite{a9}:

1. the vertex is $\gamma_{4}$;

2. the fermion line is $T\sum_{k}\int\frac{d^3p}{(2\pi)^3}G(\widetilde{p}_k)$;

3. $-Tr$ for each closed fermion loop;

4. $\beta(2\pi)^3\delta(\overrightarrow{p}_{in}-\overrightarrow{p}_{out})\delta_{\omega_{in},\omega_{out}}$ for each vertex, corresponding to energy-momentum conservation.
And $\beta(2\pi)^3\delta(0)=\beta V$.

\begin{figure}
\centering
\includegraphics[width=0.5\linewidth]{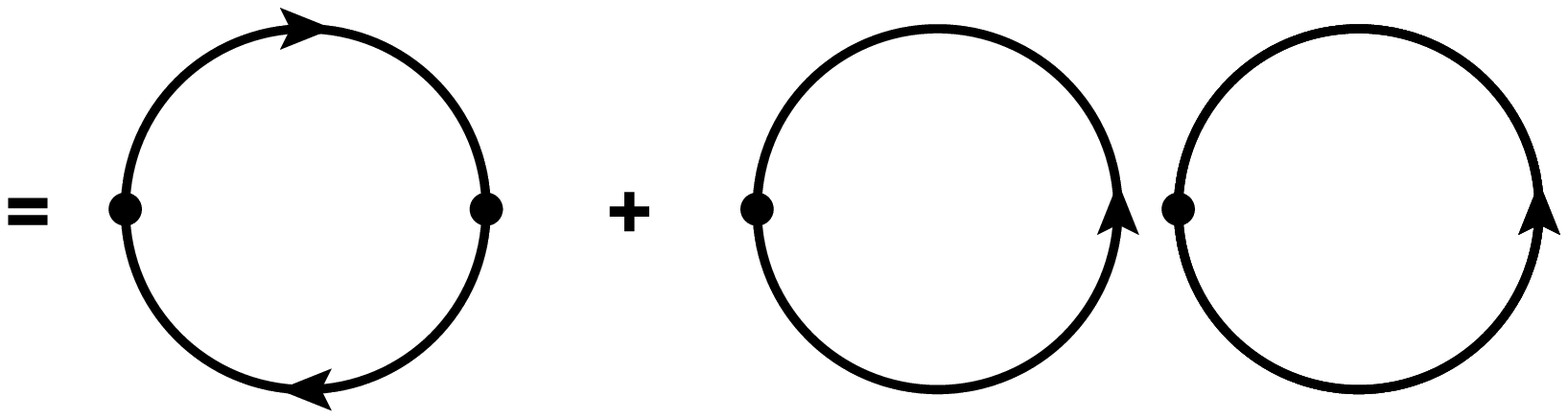}
\caption{Feynman diagram for $<N^2>$. The point represents the vertex and the line represents the quark propagator.}
\label{n2}
\end{figure}

\begin{figure}
\centering
\includegraphics[width=.8\linewidth]{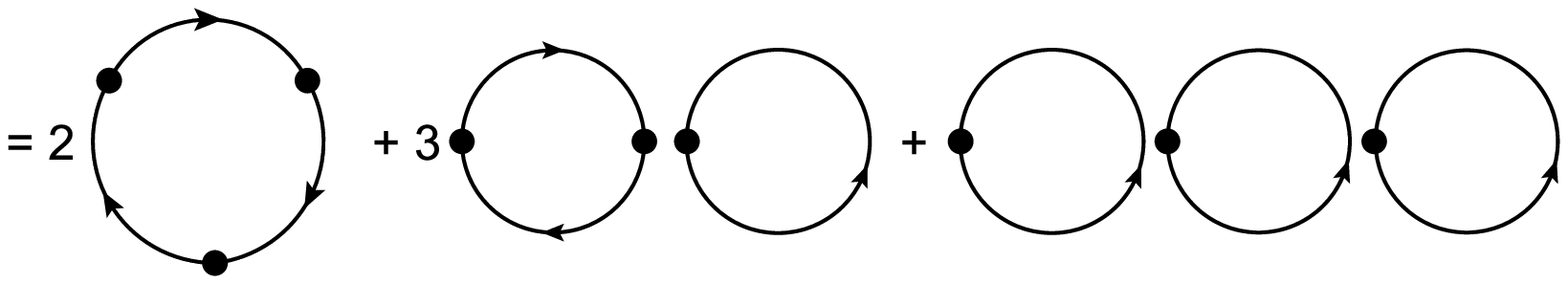}
\caption{Feynman diagram for $<N^3>$. The point represents the vertex and the line represents the quark propagator.}
\label{n33}
\end{figure}

Similarly, the expectation of $N^2$ can be expressed as
\begin{eqnarray}
\begin{split}
<N^{2}>=\frac{\int\mathcal{D}\overline{\psi}_{q}\int\mathcal{D}\psi_{q}\int_{0}^{\beta}d\tau_{1}\int d^3x_1 \overline{\psi}_{q1}\gamma_{4}\psi_{q1}\int_{0}^{\beta}d\tau_{2}\int d^3x_2 \overline{\psi}_{q2}\gamma_{4}\psi_{q2}\exp{(-\beta(H-\mu N))}}{\beta ^2*\int\mathcal{D}\overline{\psi}_{q}\int\mathcal{D}\psi_{q}\exp{(-\beta(H-\mu N))}}\ ,\\
\end{split}
\end{eqnarray}
and the Feynman diagram for $\frac{\int\mathcal{D}\overline{\psi}_{q}\int\mathcal{D}\psi_{q}\int_{0}^{\beta}d\tau_{1}\int d^3x_1 \overline{\psi}_{q1}\gamma_{4}\psi_{q1}\int_{0}^{\beta}d\tau_{2}\int d^3x_2 \overline{\psi}_{q2}\gamma_{4}\psi_{q2}\exp{(-\beta(H-\mu N))}}{(\int\mathcal{D}\overline{\psi}_{q}\int\mathcal{D}\psi_{q}\exp{(-\beta(H-\mu N))}}$ is shown in Fig. \ref{n2}. Then the variance of $N$ is
\begin{eqnarray}
\begin{split}
<(N-<N>)^2>&=<N^2>-<N>^2\\
&=-VT^2\sum_{k=-\infty}^{+\infty}\int\frac{d^3p}{(2\pi)^3}Tr[G(\widetilde{p}_k)\gamma_4G(\widetilde{p}_k)\gamma_4]\\
&=-VT^2\sum_{k=-\infty}^{+\infty}\int\frac{d^3p}{(2\pi)^3}\frac{4(\overrightarrow{p}^2+m^2-\widetilde{\omega}^2_{k})}{(\overrightarrow{p}^2+m^2+\widetilde{\omega}^2_{k})^2}\ .\\
\label{sigma}
\end{split}
\end{eqnarray}
The Feynman diagram for $<N^3>$ is shown in Fig. \ref{n33} and the third moment of $N$ is
\begin{eqnarray}
\begin{split}
<(N-<N>)^3>&=<N^3>-3<N><N^2>+2<N>^3\\
&=-2VT^3\sum_{k=-\infty}^{+\infty}\int\frac{d^3p}{(2\pi)^3}Tr[G(\widetilde{p}_k)\gamma_4G(\widetilde{p}_k)\gamma_4G(\widetilde{p}_k)\gamma_4]\\
&=2VT^3\sum_{k=-\infty}^{+\infty}\int\frac{d^3p}{(2\pi)^3}\frac{4i\widetilde{\omega}_{k}(3\overrightarrow{p}^2+3m^2-\widetilde{\omega}^2_{k})}{(\overrightarrow{p}^2+m^2+\widetilde{\omega}^2_{k})^3}\ .\\
\label{n3}
\end{split}
\end{eqnarray}

\begin{figure}
\centering
\includegraphics[width=.8\linewidth]{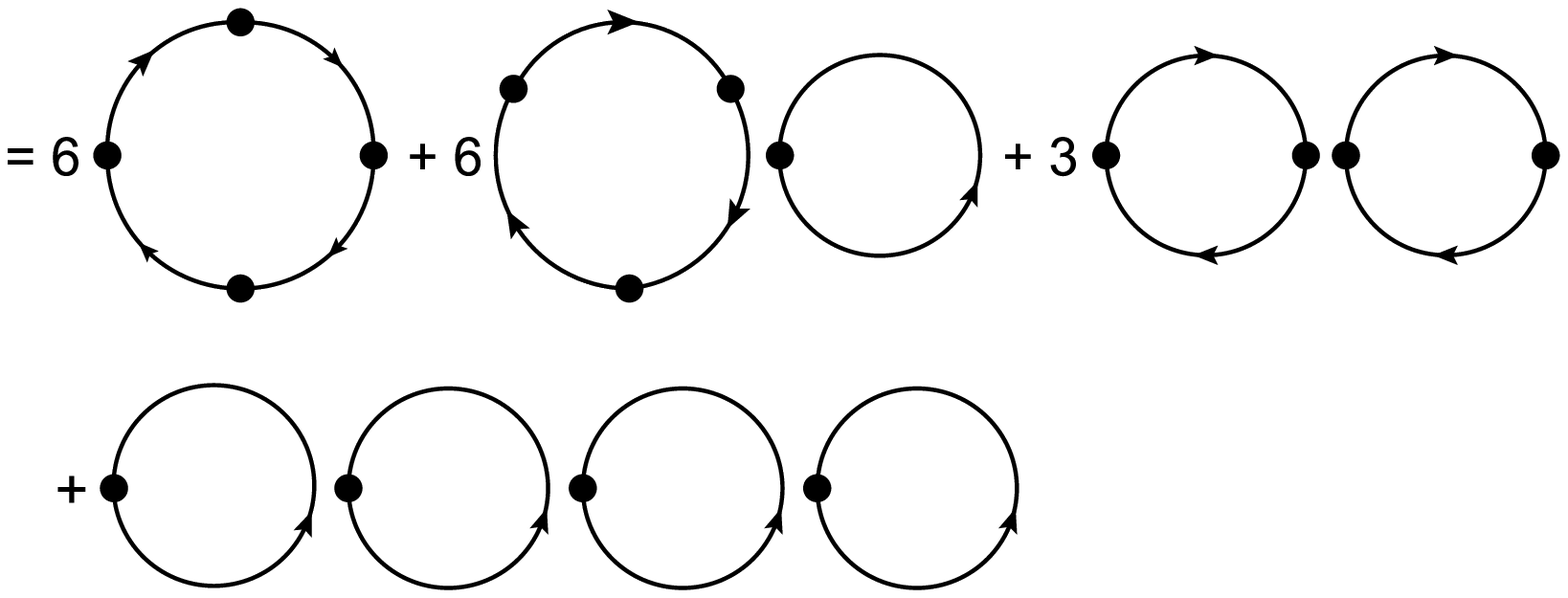}
\caption{Feynman diagram for $<N^4>$. The point represents the vertex and the line represents the quark propagator.}
\label{n44}
\end{figure}

The Feynman diagram for $<N^4>$ is shown in Fig. \ref{n44} and the numerator of $\kappa\sigma^2$ is
\begin{eqnarray}
\begin{split}
<(N-<N>)^4>&-3<(N-<N>)^2>^2\\
&=-6VT^4\sum_{k=-\infty}^{+\infty}\int\frac{d^3p}{(2\pi)^3}Tr[G(\widetilde{p}_k)\gamma_4G(\widetilde{p}_k)\gamma_4G(\widetilde{p}_k)\gamma_4G(\widetilde{p}_k)\gamma_4]\\
&=-6VT^4\sum_{k=-\infty}^{+\infty}\int\frac{d^3p}{(2\pi)^3}\frac{4((\overrightarrow{p}^2+m^2)^2-6\widetilde{\omega}^2_{k}(\overrightarrow{p}^2+m^2)+\widetilde{\omega}^4_{k})}{(\overrightarrow{p}^2+m^2+\widetilde{\omega}^2_{k})^4}\ .\\
\label{n4}
\end{split}
\end{eqnarray}

Put the results of Eq. (\ref{sigma}) and Eq. (\ref{n3}) into Eq. (\ref{s}), we can get the value of $S\sigma$,
 \begin{equation}
S\sigma=\frac{2T\sum_{k=-\infty}^{+\infty}\int\frac{d^3p}{(2\pi)^3}Tr[G(\widetilde{p}_k)\gamma_4G(\widetilde{p}_k)\gamma_4G(\widetilde{p}_k)\gamma_4]}{\sum_{k=-\infty}^{+\infty}\int\frac{d^3p}{(2\pi)^3}Tr[G(\widetilde{p}_k)\gamma_4G(\widetilde{p}_k)\gamma_4]}\ ,
\label{sss}
\end{equation}
 and put the results of Eq. (\ref{n3}) and Eq. (\ref{n4}) into Eq. (\ref{kappa}), we can get $\kappa\sigma^2$,
  \begin{equation}
\kappa\sigma^2=\frac{3T^2\sum_{k=-\infty}^{+\infty}\int\frac{d^3p}{(2\pi)^3}Tr[G(\widetilde{p}_k)\gamma_4G(\widetilde{p}_k)\gamma_4G(\widetilde{p}_k)\gamma_4G(\widetilde{p}_k)\gamma_4]}{\sum_{k=-\infty}^{+\infty}\int\frac{d^3p}{(2\pi)^3}Tr[G(\widetilde{p}_k)\gamma_4G(\widetilde{p}_k)\gamma_4]}\ .
\label{sss}
\end{equation}

\section{Results}
\label{three}

\begin{figure}
\centering
\includegraphics[width=1\linewidth]{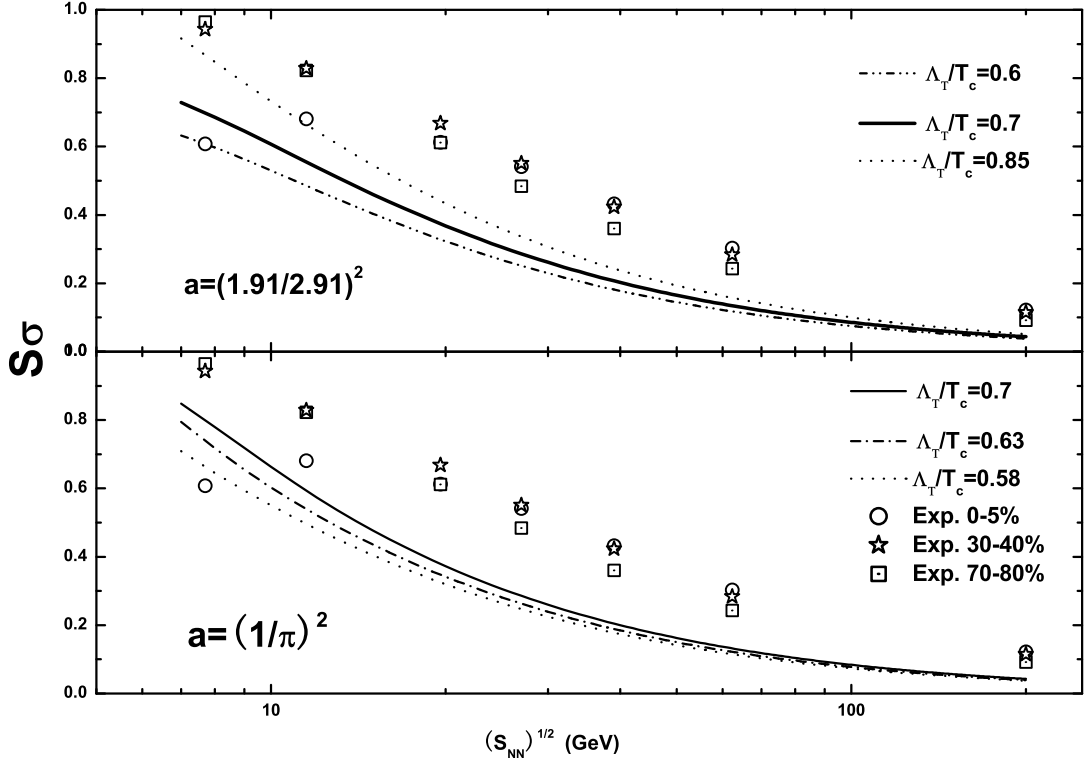}
\caption{Top three lines are $S\sigma$ results from Eq. (\ref{s}) as a function of $\sqrt{s_\mathrm{NN}}$ for $\Lambda_T/T_c=0.85,0.7,0.6$ at $a=(1.91/2.91)^2$. Bottom three lines are $S\sigma$ for $\Lambda_T/T_c=0.7,0.63,0.58$ at $a=(1/\pi)^2$. Data points from Ref. \cite{a49} are $S\sigma$ results for $0-5\%$, $30-40\%$ and $70-80\%$ Au+Au collisions respectively. }
\label{ss}
\end{figure}
\begin{figure}
\centering
\includegraphics[width=1.0\linewidth]{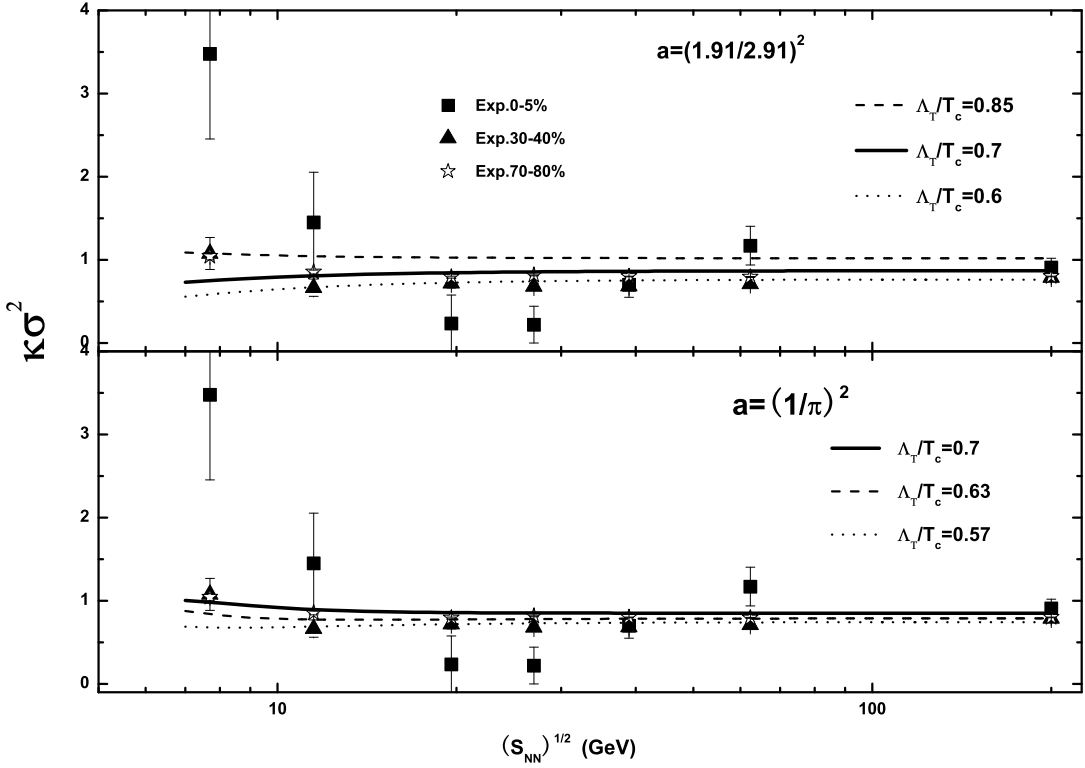}
\caption{Top three lines are $\kappa\sigma^2$ results from Eq. (\ref{kappa}) as a function of $\sqrt{s_\mathrm{NN}}$ for $\Lambda_T/T_c=0.85,0.7,0.6$ at $a=(1.91/2.91)^2$. Bottom three lines are $\kappa\sigma^2$ for $\Lambda_T/T_c=0.7,0.63,0.57$ at $a=(1/\pi)^2$. Data points from Ref. \cite{a49} are $\kappa\sigma^2$ results of Au+Au collisions at different centrality bins.}
\label{ks}
\end{figure}

The experimental results for the $S\sigma$ and $\kappa\sigma^2$ of net-proton multiplicity distributions are shown in Fig. \ref{ss} and Fig. \ref{ks} respectively. In Fig.\ref{ss}, top three lines are $S\sigma$ results from Eq. (\ref{s}) as a function of $\sqrt{s_\mathrm{NN}}$ for $\Lambda_T/T_c=0.85,0.7,0.6$ at $a=(1.91/2.91)^2$ and bottom three lines are for $\Lambda_T/T_c=0.7,0.63,0.58$ at $a=(1/\pi)^2$, where $T_c=175$ MeV is from Ref. \cite{a3}. The temperature and baryon chemical potential parameters for each energy are determined from the chemical freeze-out parameterization in heavy-ion collisions \cite{a10}. Data points are the experimental results of $S\sigma$ from Ref. \cite{a49}. In Fig. \ref{ks}, top three lines are $\kappa\sigma^2$ results from Eq. (\ref{kappa}) as a function of $\sqrt{s_\mathrm{NN}}$ for $\Lambda_T/T_c=0.85,0.7,0.6$ at $a=(1.91/2.91)^2$ and bottom three lines are for $\Lambda_T/T_c=0.7,0.63,0.57$ at $a=(1/\pi)^2$. Data points from Ref. \cite{a49} are $\kappa\sigma^2$ results of Au+Au collisions at different centrality bins.

There are two parameters $a$ and $\Lambda_T$ in our calculation. The parameter $a$ is introduced to take account of finite quark chemical potential $\mu$ \cite{a6,a7}. As mentioned above, there are mainly two choice: $a=(1.91/2.91)^2$ \cite{a6} and $a=(1/\pi)^2$ \cite{a7}. The $\Lambda_T$ is related to the QCD scale parameter. Since the second order quark number susceptibility is studied at $\Lambda_T/T_c=0.7$ \cite{a8}, the $S\sigma$ and $\kappa\sigma^2$ are calculated with $\Lambda_T/T_c$ around $0.7$. When the parameter $a$ is fixed, the values of $S\sigma$ and $\kappa\sigma^2$ are reduced with the reduction of $\Lambda_T$. In Fig. \ref{ss} the difference between the results of $S\sigma$ calculated from different $\Lambda_T$ are smaller at high energies than low energies. And for $\kappa\sigma^2$ in Fig. \ref{ks} the results at different $\Lambda_T$ are almost parallel with each other at large $\sqrt{s_\mathrm{NN}}$ and are with bigger discrepancies at small $\sqrt{s_\mathrm{NN}}$.

Particularly, the results for $\Lambda_T/T_c=0.7$ are shown as solid lines in Fig. \ref{ss} and Fig.\ref{ks}. Comparing the two solid lines in Fig. \ref{ss}, we find that the values with $a=(1.91/2.91)^2$ are lower than the one with $a=(1/\pi)^2$ at small $\sqrt{s_\mathrm{NN}}$ and the difference get smaller and smaller with increasing $\sqrt{s_\mathrm{NN}}$. As for $\kappa\sigma^2$  in Fig. \ref{ks}, at small$\sqrt{s_\mathrm{NN}}$ the two lines have different trends. The one with $a=(1.91/2.91)^2$ increases with increasing $\sqrt{s_\mathrm{NN}}$ and the other one shows opposite trend. The results of qQGP model is more sensitive to the parameters at small colliding energies.

The experimental results in Fig. \ref{ss} and Fig. \ref{ks} demonstrate that both $S\sigma$ and $\kappa\sigma^2$ clearly show non-monotonic variation for $0-5\%$ centrality when $\sqrt{s_\mathrm{NN}}$ is below 30 GeV. Above 30 GeV the results of different centrality are close to each other. The experimental results may indicate that the corresponding chemical freeze-out T and $\mu$ around 20 GeV may be close to the critical point \cite{a49}. In Fig. \ref{ss} and Fig. \ref{ks}, it is shown that our results with different parameters have the similar trends with the experimental data for the colliding energies above 30 GeV. In this region, for $S\sigma$ our results is approximately $0.1$ less than the experimental data at the maximum deviation, and for $\kappa\sigma^2$ our results can describe the experimental data well. But below 30 GeV, our results have significant discrepancies from the experimental data of $0-5\%$ centrality.

\section{Summary}
\label{four}

Baryon number fluctuations are sensitive to the QCD phase transition and QCD critical point. We calculated various order moments of the baryon number distributions in the quasi-particle model of QGP. To avoid extra undetermined term in calculating susceptibilities in quasi-particle model we try to directly calculate the various order of moments of quark number distributions. Since the term of quark number in Lagrangian is analogized as the interaction term, we can obtain the moments of quark number based on the Feynman rules of finite-temperature field theory. Finally, we compare our calculations with the latest experimental data. It is found that the results of qQGP model are more sensitive to the parameters at small colliding energies. For energies above 30 GeV, our results with different parameters have the similar trends as the experimental data. We found that the $S\sigma$ are smaller than the experimental data while the $\kappa\sigma^2$ fits the experimental data well. However, at energies below 30 GeV, our results have large discrepancies from the experimental data of $0-5\%$ centrality. These comparisons suggest that at low energies, the experimental data may contain other physics effects, for. eg. the critical point, which is not included in the qQGP model. It also indicates that the future low energy heavy-ion collisions experiment is much more important for the QCD critical point search.

\section*{Acknowledgements}

This work was supported by the MoST of
China 973-Project No. 2015CB856901 and the National Natural Science Foundation of China (under Grants No. 11447121, 11575069, 11475085, 11690030, and No. 11535005).

\end{document}